\documentclass[showpacs,twocolumn,preprintnumbers,doublespace]{revtex4}
\usepackage[dvips]{graphics}
\usepackage{amsmath, amsthm, amssymb}
\usepackage{graphicx}
\usepackage{dcolumn}
\usepackage{bm}

\newcommand{\de}[1]{\left( #1 \right)}

\newcommand{\DE}[1]{\left\{ #1 \right\}}

\newcommand{\ket}[1]{\left| #1 \right\rangle}
\newcommand{\bra}[1]{\left\langle #1 \right|}

\newcommand{\mean}[1]{\left\langle #1 \right\rangle}

\newcommand{\tr}{\mathrm{Tr}}

\newcommand{\eg}{{\it{e.g.}}, }
\newcommand{\ie}{{\it{i.e.}}, }

\usepackage{dcolumn}
\usepackage{bm}

\begin{document}

\title{Estimating entanglement of unknown states}

\author{Daniel Cavalcanti}
\email{dcs@fisica.ufmg.br} \affiliation{Departamento de F\'{\i}sica
- Caixa Postal 702 - Universidade Federal de Minas Gerais -
30123-970 - Belo Horizonte - MG - Brazil}

\author{Marcelo O. Terra Cunha}
\email{tcunha@mat.ufmg.br} \affiliation{Departamento de Matem\'atica
- Caixa Postal 702 - Universidade Federal de Minas Gerais -
30123-970 - Belo Horizonte - MG - Brazil} \affiliation{The School of
Physics and Astronomy, University of Leeds, Leeds LS2 9JT, UK}

\begin{abstract}
The experimental determination of entanglement is a major goal in
the quantum information field. In general the knowledge of the state
is required in order to quantify its entanglement. Here we express a
lower bound to the robustness of entanglement of a state based only
on the measurement of the energy observable and on the calculation
of a separability energy. This allows the estimation of entanglement
dismissing the knowledge of the state in question.
\end{abstract}
\maketitle

Entanglement - purely quantum correlations - has been viewed as the
main resource that allows several practical tasks such as quantum
cryptography \cite{ekert}, teleportation \cite{telep}, and quantum
computation \cite{comp}. Despite the great efforts that have
resulted in important advances in the area of Quantum Information,
this resource is still to be completely understood. One important
question is how to properly quantify entanglement and how to measure
the amount of entanglement present in a system in real experiments,
especially in the multipartite scenario.

One of the difficulties in dealing with the quantification of
entanglement is the fact that most of the proposed quantifiers use
mathematical operations without clear physical interpretations.
Furthermore an extra difficulty appears when we do not know
precisely  the state of the system we are dealing with, since the
majority of the quantifiers are based on the knowledge of the
density matrix of the system\footnote{This assumption is very
natural from a theoretical point of view. However, in practical
applications, it implies the necessity of a tomographic-like process
to completely determine the state, prior to quantify its
entanglement.}.

One alternative approach is just to detect entanglement, without
quantifying it. A powerful method that can be used for this aim is
the use of {\emph{entanglement witnesses}}\cite{detect}. An
entanglement witness for the entangled state $\rho$ is given by a
Hermitian operator $W$ such that
\begin{eqnarray}
&\tr(W\rho)<0,&\\
\label{sep} {\text{while}}&
\tr(W\sigma)\geq 0,&  \forall \sigma \in \mathcal{S},
\end{eqnarray}
where $\mathcal{S}$ denotes the set of separable states (\ie states with classically described correlations \cite{Wer}).
This is a simple consequence of the structure of the set of separable states: $\mathcal{S}$ is a convex closed set. An important
point about entanglement witnesses appears in the multipartite case: they can be used to detect different kinds of entanglement,
just defining $\mathcal{S}$ as the set of states which do not have such a kind of entanglement. Another important advantage is that,
as $W$ is a Hermitian operator, it can be seen as an {\emph{observable}}, and can be directly measured \cite{expW}. Through this road, one goal
was already achieved: experimental detection of entanglement without the previous determination of the quantum state.

In this Letter we aim to achieve another goal: we show a method for
estimating {\emph{how much}} entanglement a quantum state has,
without the need of knowing it. More precisely, we show how to
obtain a lower bound of the {\emph{generalized robustness}}
\cite{steiner} of an unknown state by measuring only the expected
value of an observable: energy. One must appreciate the importance
of describing relevant physical properties of the system by a small
amount of numbers (in the present case, just one), instead of a
detailed ``microscopic'' description of the system (here, the
knowledge of its state), as is done by Thermodynamics. One should
remember that even the most simple system to show entanglement (two
qubits) needs $15$ numbers to be completely
characterized\footnote{The independent 15 real parameters that
determines a density matrix. Of course, if one make additional
assumptions, as considering the global state pure, this number
lowers.}, which makes the state reconstruction highly inefficient.

Recently, it has been shown that this thermodynamical analogy can be
made much more deep. In fact, some thermodynamical properties can be
regarded as true entanglement witnesses for some systems
\cite{Bose,BVZ,PRB,TermEnt,Toth,DDB04}. Specifically, it was shown
that if the ground state of a Hamiltonian is entangled, a
measurement of energy can directly show that the system is entangled
\cite{Toth}. This can be understood by noting that in this case
there will be a lower bound to the energy of all separable states,
denoted by $E_{sep}$, which is higher than the ground state energy.
Consequently, if the system is found with less energy than $E_{sep}$
it is automatically known to be entangled. The value of $E_{sep}$
was determined for many interesting systems such as Heisenberg
spin chains \cite{Toth,DDB04} and the Bose-Hubbard model
\cite{Toth}.

Following T\'oth \cite{Toth}, one can rephrase the above discussion
in terms of entanglement witnesses. One can define the energy-based
entanglement witness $ W=H-E_{sep}I, $ where $H$ is the Hamiltonian,
$I$ is the identity operator, and $\displaystyle{
E_{sep}=\min_{\ket{\psi} \in {S}} \bra{\psi}H\ket{\psi}}, $ where
$S$ denotes the set of pure separable sates. This definition
guarantees the relation \eqref{sep} to hold and, if $E_{sep}$ is
different from the ground state energy, all states with
$\left\langle E\right\rangle < E_{sep}$ will be detected by $W$. In
this case, one calls $E_{sep} - E_0$ the {\emph{entanglement gap}}
($E_0$ being the ground state energy), as it gives the range in
energy of the states which entanglement is unveiled by such a
witness \cite{DDB04}.

Up to this point, entanglement witnesses were said to only detect
entanglement. However, they can also be used in {\emph{quantifying}}
it as well \cite{BNT02,BV04,Fernando2}. Following
ref.~\cite{Fernando2}, one can define a {\emph{witnessed
entanglement}} of the state $\rho$ as
\begin{equation}\label{WE}
E\de{\rho}=\max\DE{0,-\min_{W\in\mathcal{M}}Tr(W\rho)},
\end{equation}
where $\mathcal{M}$ is a restricted set of entanglement witnesses.
Depending on this restriction, different entanglement quantifiers
appear (including some well known ones as special cases). In this
Letter we shall only deal with the {\emph{generalized robustness}}
\cite{steiner}, which corresponds to the case with $\mathcal{M}$
given by the restriction $W \leq I$.

The generalized robustness of $\rho$ ($R_g (\rho)$) is defined as
the minimum amount of mixing with another state (by means of a
convex combination) such that this mixture losses its entanglement.
More precisely, it is given by the minimum $s$ such that $\frac{\rho
+ s\pi}{1+s}$ is a separable state, with $\pi$ representing any (not
necessarily separable) state. It was found to have a direct
information theoretical interpretation as the best fidelity of
teleportation one can reach by using $\rho$ as a quantum channel (in
the 2-qubit case) \cite{Verstraete}, or  the improvement $\rho$
causes to the teleportation process if it is used as an ancillary
state (for general bipartite states) \cite{Fernando}. It is thus
important to estimate this entanglement quantifier from an
operational point of view. We now set a scheme to obtain a lower
bound for the generalized robustness of an unknown state.

For an entangled state $\rho$, the above discussion implies
\begin{equation}
R_{g}\de{\rho}=-\tr\de{W_{opt}\rho}= -\left\langle
W_{opt}\right\rangle ,
\end{equation}
where $W_{opt}$ is an optimal entanglement witness for $\rho$ in the sense of the minimization procedure in Eq.~\eqref{WE}.
The crucial point is that for an unknown state one can not determine an optimal witness.

Suppose now we are dealing with a bounded Hamiltonian $H$ (\eg a
finite dimensional Hilbert space). Set
$A = \sup \left| \left\langle H\right\rangle - E_{sep}\right|,
$ where the suppremum is taken over all quantum states, and define
\begin{equation}\label{W}
W=\frac{H-E_{sep}I}{A}. \end{equation} This $W$ is an entanglement
witness for all states with $\mean{H} < E_{sep}$ and obeys $W \leq
I$. Hence, if $\mean{H} < E_{sep}$, we have
\begin{equation}\label{E}
R_g\de{\rho} \geq \frac{E_{sep} - \langle H\rangle}{A},
\end{equation}
since one can not guarantee that this witness is optimal.

In the multipartite scenario, the situation is even better: the
measurement of one observable, the energy, can estimate various
kinds of entanglement. For this, one only needs to be able to
calculate the various values of $E_{sep}$, one for each kind of
entanglement. Whenever $E_{sep}$ is greater than the measured
energy, Eq.~\eqref{E} gives a lower bound for that kind of
entanglement. The quantity $E_{sep}$ can be obtained by the
techniques in Refs.~\cite{DDB04}. Naturally, this scheme works
better for states with low energy, like low temperature thermal
equilibrium states. However, it is also important that the scheme
can be used for any state, with the only usual condition of
reproducibility: \ie one must prepare a whole ensemble of copies
characterized by $\rho$ to get a good evaluation of $\mean{E}$, and
from this obtain the estimates of entanglement.

\begin{figure}\centering
   \rotatebox{270}{\includegraphics[width=6cm]{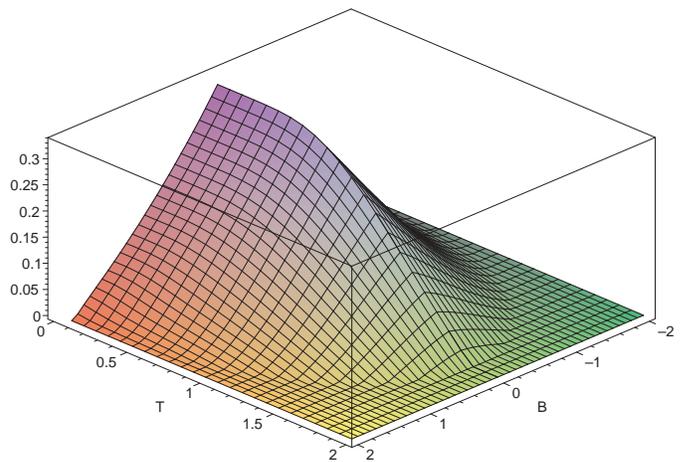}}
    \caption{(Color online) Lower bound to $R_g(\rho)$ given by Eq.~\eqref{E}
    to the state of thermal equilibrium of the Hamiltonian $H_{XXX}$ (we used $k_{B}=1$ and         $J=1$).}
    \label{Lb}
\end{figure}

As a matter of illustration we exemplify our results using a
two-qubit system coupled through a Heisenberg $XXX$ interaction,
with coupling constant $J > 0$, subjected to a magnetic field $B$:
\begin{equation}
H_{XXX} = J \vec{\sigma _1}\cdot \vec{\sigma _2} + B \de{\sigma _{z1} + \sigma _{z2}},
\end{equation}
using $E_{sep}$ as calculated in Ref.~\cite{Toth}. In Figure
\ref{Lb} we have displayed the lower bound \eqref{E} for the state
of thermal equilibrium $\rho(T)=Z^{-1}{\exp(-\beta H)}$, where
$Z=\tr \exp(-\beta H)$ is the partition function and $\beta=(k_B
T)^{-1}$, $k_B$ denoting the Boltzmann constant and $T$ the absolute
temperature. The behavior at the picture is qualitatively in
accordance with our intuition. For $B=0$ the entanglement is
greater, but even in this case, for a finite temperature it will
vanish, in an entangled-disentangled transition \cite{QPT}. As Eq.~\eqref{E} only gives a lower bound to
entanglement, the precise situation is a little bit different. For
example, for $B=0$ and $T=0$, we have nothing more than the singlet,
and its generalized robustness is $1$, despite the energy-based
witness only imply $R_g \geq 0.33$. Also the transition truly occurs
at $T_c = 3.65$, while here the estimate only says that $T_c \geq
1.82$.

It is important to stress that the critical temperature $T_{sep}$
below which the system is certainly entangled was already estimated
for real systems and can give a notion of the limits within our
approach works. In Ref.~\cite{BVZ} the authors found
$T_{sep}\approx5.6K$ for the cooper nitrate ($\mathrm{CN}$).
Although it is expected that our method works better for low
temperatures, some systems can show a separability temperature as
high as $T_{sep}\approx365K$, as is the case of the nanotubular
system $\mathrm{Na_2 V_3 O_7}$ \cite{PRB}!

Some related previous work deserves mention. Ref.~\cite{HorEk} shows
how to implement experimentally maps related to positive but not
completely positive maps on $n$ copies of a given state. This
strategy permits the detection of entanglement in the bipartite
scenario. Ref.~\cite{Steve} reports the measurement of concurrence
on hyperentangled states, in which two logical copies of a qubit are
encoded in just one physical system. Our approach works in the
so-called one-copy regime, which usually is simpler from the
experimental point of view, and is directly applicable for the
estimation of multipartite entanglement as well.

Summarizing, we have presented a way of estimating the entanglement
of a system without having previous knowledge of it. This method
relies on a lower bound to the generalized robustness of
entanglement which is given through the measurement of the mean
value of the energy only. This special entanglement quantifier has
an operational interpretation in terms of the best fidelity of
teleportation. It is important to emphasize the advantage of
measuring, or at least estimating, entanglement quantifiers for
practical applications. We hope the present discussion
help in this task and also add flavor on the experimental
quantification of entanglement.

The authors  thank CNPq and PRPq-UFMG for financial
support. This work is part of the Brazilian Millennium Institute for
Quantum Information. The authors also thank Marcelo Fran\c ca Santos,
Leandro Malard, and Libby Heaney for enlightening discussions on
this theme.

\end{document}